\documentclass[preprint,aps,12pt,preprintnumbers,nofootinbib,superscriptaddress] {revtex4}

\pdfoutput=1

\usepackage{graphicx}
\usepackage{amssymb}
\usepackage{amsmath}
\usepackage{epstopdf}
\usepackage{subfigure}
\usepackage{setspace}

\def\be{\begin{equation}}
\def\ee{\end{equation}}
\def\bea{\begin{eqnarray}}
\def\eea{\end{eqnarray}}

\def\wb{\bar{w}}
\def\Xb{\bar{X}}

\def\vp{\hat{p}}

\preprint{IHES/P/12/03}

\begin{document}

\title{Algebras for Amplitudes}

\author{N.~E.~J.~Bjerrum-Bohr}
\affiliation{Niels Bohr International Academy and Discovery Center, Niels Bohr Institute, University of Copenhagen, Blegdamsvej 17, DK-2100 Copenhagen, Denmark}
\affiliation{Institut des Hautes \'Etudes Scientifiques, 91440, Bures-sur-Yvette, France}
\author{Poul~H.~Damgaard}
\author{Ricardo~Monteiro}
\author{Donal~O'Connell}
\affiliation{Niels Bohr International Academy and Discovery Center, Niels Bohr Institute, University of Copenhagen, Blegdamsvej 17, DK-2100 Copenhagen, Denmark}

\date{\today \\ \phantom{a}}

\begin{abstract}
Tree-level amplitudes of gauge theories are expressed in a basis of auxiliary amplitudes with only cubic vertices. The vertices in this formalism are explicitly factorized in color and kinematics, clarifying the color-kinematics duality in gauge theory amplitudes. The basis is constructed making use of the KK and BCJ relations, thereby showing precisely how these relations underlie the color-kinematics duality. We express gravity amplitudes in terms of a related basis of color-dressed gauge theory amplitudes, with basis coefficients which are permutation symmetric.
\end{abstract}

\maketitle

\section{Introduction}

\noindent
In recent years, our knowledge and understanding of scattering amplitudes in our most cherished theories has been revolutionised. Of course, the present interest in scattering amplitudes was spurred by our need to compute cross-sections at the LHC. But there has been much progress in areas of research which are not obviously related to the LHC. The topic of this article is one such area; namely, the correspondence between scattering amplitudes in gauge theory and in gravity. 

Indeed, one of the promises of our recent advances is that the fascinating connection between gauge theory and gravity amplitudes may be illuminated. The intricate KLT relations between tree-level scattering amplitudes in Yang-Mills theory and gravity were outlined long ago~\cite{Kawai:1985xq}. Recently, progress in conveniently expressing  gravity amplitudes as the square (or double-copy) of gauge theory amplitudes~\cite{Bern:2008qj,Bern:2010yg}, including at loop level~\cite{Bern:2010ue}, has allowed for remarkable studies of the ultraviolet properties of supergravity theories~\cite{Bern:2010tq,Carrasco:2011mn,Bern:2011rj,Naculich:2011fw,BoucherVeronneau:2011qv,Bern:2012uf,Bern:2012cd}. This new squaring relation between gravity and gauge theory relies on an unexpected property of gauge theory amplitudes: the existence of a duality between color and kinematics, when the amplitude is appropriately written in terms of diagrams with only cubic vertices. This duality hints at a Lie algebra structure underlying the kinematic dependence of the amplitudes, akin to the Yang-Mills Lie algebra. Such a kinematic Lie algebra was indeed recently found for MHV amplitudes~\cite{Monteiro:2011pc}, where it was shown to be inherited from the symmetries of the self-dual sector. Generally, however, there should exist a structure independent of helicities or spacetime dimensions. 

In this article, we explore how to write gluon tree-level amplitudes in gauge theory using 
cubic vertices only, thus making manifest the intriguing duality between color and kinematics. Through the squaring relation, we draw conclusions for gravity amplitudes. Although we consider gluon amplitudes for explicit checks, these results are valid also for supersymmetric extensions with fermions and scalars.

We reverse the original reasoning of the color-kinematics duality~\cite{Bern:2008qj}. In the latter work, it was found that scattering amplitudes can be written such that the color-kinematics duality is satisfied, and that this implies the existence of certain linear relations between color-ordered amplitudes, known as the BCJ relations. These relations hold in addition to the previously known KK relations~\cite{Kleiss:1988ne}. The KK and BCJ relations have been proven as identities based on monodromy in string theory amplitudes~\cite{BjerrumBohr:2009rd,Stieberger:2009hq}, and also directly in field theory using the BCFW recursion method \cite{Feng:2010my}. So the main question that we address here is how the KK and BCJ relations imply a Lie algebra structure for the kinematic dependence of the amplitudes.

This paper is organized as follows. In Section~\ref{sec:review}, we review the color-kinematics duality. In Section~\ref{sec:cubicScalar}, we construct simple theories whose amplitudes satisfy this duality. In Section~\ref{sec:cubicYM}, we use these theories to construct a basis for gauge theory amplitudes, making manifest the color-kinematics duality. This is our main result. In Section~\ref{sec:colordual}, we study a certain decomposition of gauge theory amplitudes suggested by the duality. In Section~\ref{sec:gravity}, we comment on implications for gravity amplitudes. Finally, in Section~\ref{sec:concl}, we present our conclusions.

\section{Review}
\label{sec:review}

A tree-level gauge theory scattering amplitude ${\mathcal A}$ can be expanded as 
\be
\label{standarddecA}
{\mathcal A} = g^{n-2}  \sum_\sigma  \textrm{Tr}[T^{a_1}T^{a_2}\cdots T^{a_n}] \, A(1,2,\ldots,n)\,,
\ee
where the $T^{a}$'s denote the fundamental representation matrices of the Lie gauge group, and the sum is over non-cyclic permutations of external legs\footnote{In  Eq.~\eqref{standarddecA} we suppress the action of the permutation $\sigma$ on the particle numbers $2, 3, \ldots n$ for notational simplicity. We adopt this convention throughout the article.}. The gauge invariant components $A(1,2,\ldots,n)$ are referred to as {\it color-ordered amplitudes}. The existence of linear relations between the color-ordered amplitudes, however, means that we can do better than~\eqref{standarddecA}. The KK relations~\cite{Kleiss:1988ne} allow us to express all color-ordered amplitudes in terms of a set of $(n-2)!$ of them. The recently found BCJ relations~\cite{Bern:2008qj} further reduce the number of independent color-ordered amplitudes to $(n-3)!$. Both the KK and the BCJ relations have been proven for any number of external legs~\cite{BjerrumBohr:2009rd,Stieberger:2009hq,Feng:2010my}. A set of $(n-3)!$ independent color-ordered amplitudes can be chosen by considering permutations of all but three external legs.

The BCJ relations were obtained in \cite{Bern:2008qj} as a result of a surprising property of gauge theory amplitudes:
that they can always be written as a sum over diagrams with only cubic vertices, the cubic structure being determined by the color factors appearing at each vertex:
\begin{equation}
\label{bcjrep}
\mathcal{A} = g^{n-2}  \sum_i \frac{n_i \, c_i}{D_i},
\end{equation}
where $c_i$ are the colour factors, $D_i$ are the products of Feynman propagators, and $n_i$ are the numerators of those diagrams, which carry information about momenta and helicities. The contribution of a four-point vertex in a diagram can always be written in terms of two three-point vertices, by introducing a propagator and multiplying the numerator of the diagram by the corresponding factor. There is great freedom in choosing the kinematic numerators $n_i$, since this procedure is not uniquely determined. A choice which has proven very insightful (and which does not fix this freedom completely) is to demand that these kinematic numerators satisfy identities analogous to the Jacobi identities satisfied by the color factors $c_i$,
\begin{equation}
c_i \pm c_j \pm c_k = 0    \qquad \longrightarrow \qquad   n_i \pm n_j \pm n_k = 0.
\end{equation}
There is then a duality between color and kinematics. The fact that this representation is possible for gauge theory amplitudes (and for amplitudes of the closely related theories to be discussed next) implies linear relations among color-ordered amplitudes, the BCJ relations. The color-kinematics duality has been discussed extensively in recent work, see~\cite{Sondergaard:2009za,BjerrumBohr:2010zs,Tye:2010dd,Jia:2010nz,BjerrumBohr:2010ta,Tye:2010kg,Vaman:2010ez,Chen:2010ct,BjerrumBohr:2011kc,Bern:2011ia,BjerrumBohr:2011xe,Mafra:2011kj,Du:2011js,Sondergaard:2011iv,Broedel:2011pd,Ma:2011um,Boels:2011tp,Naculich:2011my,Jin:2012mk}.

A remarkable aspect of this representation is that it allows for a simple construction of gravity amplitudes as the ``square" of gauge theory amplitudes~\cite{Bern:2008qj,Bern:2010yg},
\begin{equation}
\label{square}
\mathcal{M} = i\left(  \frac{\kappa}{2}  \right)^{n-2}  \sum_i \frac{n_i \, \tilde{n}_i}{D_i}.
\end{equation}
Here, we allow for the fact that $n_i$ and $\tilde{n}_i$ need not correspond to the same gauge theory, {\it e.g.} for amplitudes in ${\mathcal N}=4$ supergravity, one may take $n_i$ from ${\mathcal N}=4$ super-Yang-Mills, and $\tilde{n}_i$ from pure Yang-Mills. Moreover, it is enough that only the numerators $n_i$ (or $\tilde{n}_i$) satisfy Jacobi-like identities \cite{Bern:2010yg}.

One natural puzzle arises from the duality between color and kinematics. While the color factors $c_i$ follow straightforwardly from sewing together the structure constants $f^{abc}$ for each diagram with cubic vertices, the numerators $n_i$ are not easily given a vertex interpretation. One of the obstacles is the four-point interaction vertex, which traditionally appears in formulations of gauge theory. Yet we know from BCFW recursion
that all $n$-point scattering amplitudes can be computed recursively using only knowledge of  
three-point amplitudes. This already hints at a possible special role played by three-vertices.  
An underlying three-vertex structure for $n_i$ was found in \cite{Monteiro:2011pc} for MHV amplitudes. 
That structure is inherited from the self-dual sector of the gauge theory, which will be reviewed below. 
We will use these results as a guide to understand the general case.

\section{Theories with Cubic Vertices}
\label{sec:cubicScalar}
\noindent
Let us begin by exploring simple theories which share some of the features of Yang-Mills amplitudes but which clearly have Feynman diagram expansions with only cubic vertices, and also have naturally defined numerators satisfying Jacobi identities. 

\subsection{Self-dual Yang-Mills}

Restricting to the self-dual sector of Yang-Mills theory removes the quartic term in the Lagrangian. There is only one degree of freedom left in the gauge field, which may be computed from a Lie algebra-valued scalar field $\Phi$. The equation of motion is~\cite{cubic}
\begin{equation}
\partial^2 \Phi + i g [ \partial_w \Phi, \partial_u \Phi] = 0,
\end{equation}
where we have chosen light-cone coordinates
\begin{equation}
u = t-z, \quad w = x+iy.
\end{equation}
In momentum space, the cubic vertex in this theory is given by
\vspace{-2.3cm}
\begin{equation}
\setlength{\unitlength}{2cm}
\begin{picture}(2,2)
\put(-1.5,0){\line(-1,1){0.6}}
\put(-1.5,0){\line(-1,-1){0.6}}
\put(-1.5,0){\line(1,0){0.849}}
\put(0.2,0){$=\; g\, (2 \pi)^4 \delta^{(4)}(p_1+p_2 +p_3) X(p_1, p_2) f^{abc}, $}
\put(-2.3,0.6){$1$}
\put(-2.3,-0.7){$2$}
\put(-0.5,0.0){$3$}
\end{picture}
\end{equation}
\vspace{.8cm}

\noindent
where $X(p_1, p_2)$ is a linear function of $p_1$ and $p_2$ given by
\begin{equation}
X(p_1, p_2) = p_{1w} p_{2u} - p_{1u} p_{2w}.
\end{equation}
As emphasized in~\cite{Monteiro:2011pc}, this three-point function is the product of the structure constants of two Lie algebras: obviously one of these factors is $f^{abc}$, the structure constants of the gauge group of the theory, while the other factor is
\begin{equation}
F^{p_1 p_2 p_3} =  (2 \pi)^4 \delta^{(4)}(p_1+p_2 +p_3) X(p_1, p_2),
\end{equation}
the structure constants of an area-preserving diffeomorphism Lie algebra.

The amplitudes in this theory, seen as a sector of Yang-Mills theory, correspond to helicity configurations $-+\cdots+$, and therefore vanish on-shell. However, they satisfy the KK and BCJ relations off-shell in this formalism. In the following, we consider a generalization which is non-trivial on-shell.

\subsection{A more general diffeomorphism algebra}

The self-dual Yang-Mills amplitudes are built from cubic vertices which contain an area-preserving diffeomorphism algebra. We may generalize the theory by replacing the $X(p_1, p_2)$ factor with a more general object $Y(p_1, p_2)$. Indeed, to maintain the Lie algebra structure, we only need to require that $Y(p_1, p_2)$ be linear in $p_1$ and $p_2$ and antisymmetric under exchange of $p_1$ and $p_2$. Therefore, $Y$ is specified by a choice of two-form $\Omega$:
\begin{equation}
Y(p_1, p_2) = p_{1 \mu} \, \Omega^{\mu \nu} \, p_{2 \nu} ,  \qquad   \Omega^{\mu \nu} = - \Omega^{\nu \mu}.
\end{equation}
In the case of the self-dual theory, the two-form happens to be self-dual. We choose to relax this property. In fact, we can consider the theory in arbitrary spacetime dimensions $d$. Thus, we consider a more general Lie algebra,
\be
[L_{p_1},L_{p_2}] = Y(p_1,p_2) \, L_{p_1+p_2},  \qquad L_p=i e^{-ip\cdot x} \Omega^{\mu \nu} p_\mu \partial_\nu.
\ee
The structure constants can be written as
\be
\label{defF}
F^{p_1 p_2 p_3} = (2 \pi)^d \delta^{(d)} (p_1+p_2+p_3) Y(p_1,p_2),
\ee
which is completely antisymmetric and cyclic in its indices, just as $f^{abc}$. Following \cite{Monteiro:2011pc}, we adopt an integral Einstein convention for the contraction of the indices,
\begin{align}
F^{p_1 p_2 q} \,F^{p_3 p_4}{}_q & = (2 \pi)^d\int d^d q \, \delta^{(d)}(p_1+p_2+q) \, Y(p_1,p_2) \, \delta^{(d)}(p_3+p_4-q) \, Y(p_3,p_4) \nonumber \\
& = (2 \pi)^d\delta^{(d)}(p_1+p_2+p_3+p_4) \, Y(p_1,p_2) Y(p_3,p_4) .
\end{align}
The indices are lowered and raised using $\delta^{p q} = \delta_{p q} = (2 \pi)^d \delta^{(d)}(p+q)$. The Jacobi identity takes the form
\be
F^{p_1 p_2 q} \,F^{p_3 p_4}{}_q + F^{p_2 p_3 q} \,F^{p_1 p_4}{}_q + F^{p_3 p_1 q} \,F^{p_2 p_4}{}_q = 0.
\ee

We define our theory by specifying this single cubic vertex. The resulting theory is not Lorentz invariant, but has non-trivial amplitudes which satisfy the KK and BCJ relations, since these relations follow from the Jacobi identities for the numerators.

\subsection{General Lie algebras}

It should now be clear that it is straightforward to build theories whose amplitudes satisfy the KK and BCJ relations, and which have only cubic vertices. We merely define the vertex to be the product of the structure constants of two Lie algebras,
\vspace{-2.3cm}
\begin{equation}
\label{eq:3pt}
\setlength{\unitlength}{2cm}
\begin{picture}(0,2)
\put(-1.5,0){\line(-1,1){0.6}}
\put(-1.5,0){\line(-1,-1){0.6}}
\put(-1.5,0){\line(1,0){0.849}}
\put(0.2,0){$=\quad  {\mathfrak f}^{\hat{a}\hat{b}\hat{c}} \;f^{abc}. $}
\put(-2.3,0.6){$1$}
\put(-2.3,-0.7){$2$}
\put(-0.5,0.0){$3$}
\end{picture}
\end{equation}
\vspace{.6cm}

\begin{figure}
\centering
\includegraphics[scale=0.5]{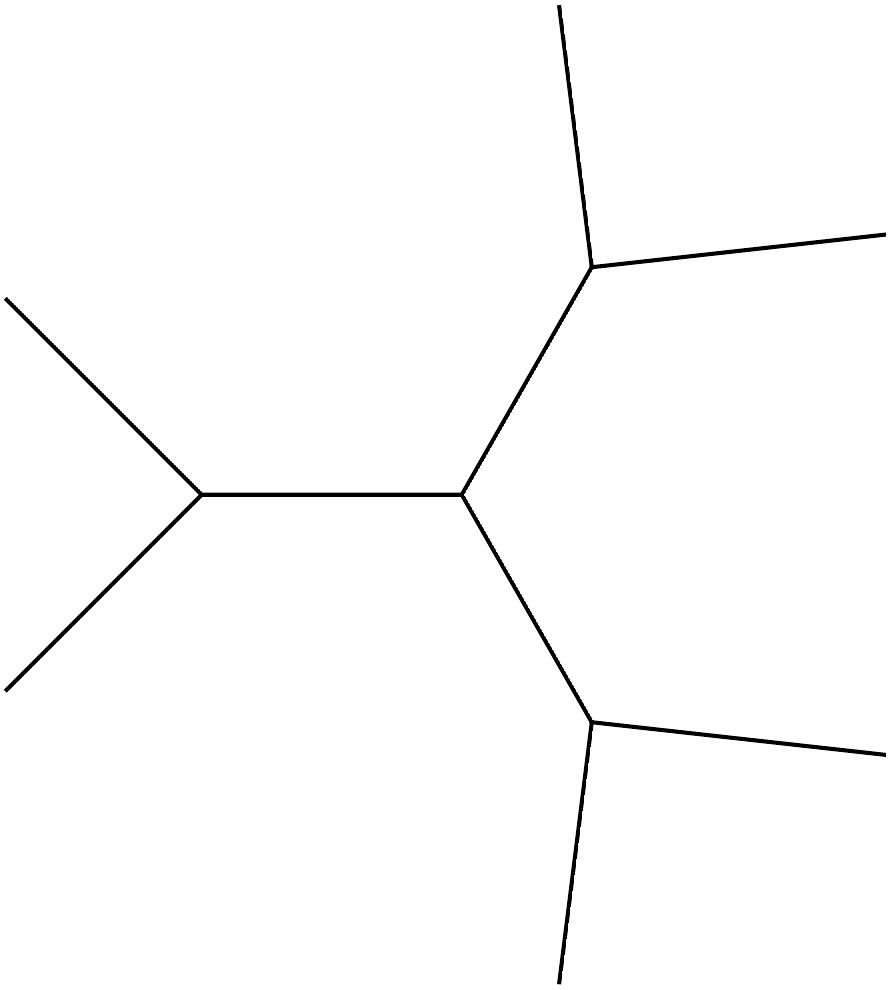}
\put(2,10){$3$}
\put(-21,-3){$6$}
\put(-21,50){$2$}
\put(-50,35){$1$}
\put(2,38){$4$}
\put(-50,14){$5$}
\caption{Example of a diagram contributing to the 6-point amplitude.}
\label{fig:6pt}
\end{figure}
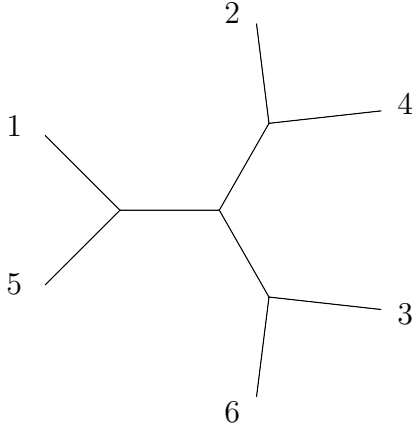

\noindent
As an example, consider the six-particle diagram in Figure~\ref{fig:6pt}.
In such a theory, its contribution to the amplitude is
\be
\label{YfD}
\frac{ \overbrace{ {\mathfrak f}^{\hat{a}_1 \hat{a}_5 \hat{b}_1}  {\mathfrak f}^{\hat{a}_2 \hat{a}_4 \hat{b}_2} {\mathfrak f}^{\hat{a}_3 \hat{a}_6 \hat{b}_3}  {\mathfrak f}^{\hat{b}_1 \hat{b}_2 \hat{b}_3} }^{n_i}\,
 \overbrace{f^{a_1 a_5 b_1}  f^{a_2 a_4 b_2} f^{a_3 a_6 b_3}  f^{b_1 b_2 b_3}  }^{c_i}  }{\underbrace{(p_1+p_5)^2(p_2+p_4)^2(p_3+p_6)^2}_{D_i}}\,.
\ee

In the case of the self-dual theory and the more general diffeomorphism theory, we chose an infinite-dimensional Lie algebra, ${\mathfrak f}^{\hat{a}\hat{b}\hat{c}} \to F^{p_1p_2p_3}$, using the particle's momenta as labels 
in the structure constants. One can also choose to use this type of Lie algebra but allocating 
fictitious momentum labels $\hat{p}_a$ arbitrarily to the participating legs 
(momentum conservation must still hold). 
Consider again the six-particle diagram in Figure~\ref{fig:6pt}. Its contribution is
\begin{align}
&\frac{ F^{\vp_1 \vp_5 q_1} \, F^{\vp_2 \vp_4 q_2} \, F^{\vp_3 \vp_6 q_3} \, F_{q_1 q_2 q_3}\,
 f^{a_1 a_5 b_1}  f^{a_2 a_4 b_2} f^{a_3 a_6 b_3}  f^{b_1 b_2 b_3}  }{(p_1+p_5)^2(p_2+p_4)^2(p_3+p_6)^2} = \nonumber \\
&\quad = \frac{ Y(\vp_1,\vp_5)  Y(\vp_2,\vp_4)  Y(\vp_3,\vp_6)  Y(\vp_1+\vp_5,\vp_2+\vp_4)  \,
 f^{a_1 a_5 b_1}  f^{a_2 a_4 b_2} f^{a_3 a_6 b_3}  f^{b_1 b_2 b_3}  }{(p_1+p_5)^2(p_2+p_4)^2(p_3+p_6)^2} \times  \nonumber \\
& \qquad \times (2\pi)^d \delta^{(d)}\left(\sum_{a=1}^6 \hat{p}_a\right),
\label{FfD}
\end{align}
where the physical momenta $p_a$ run in the propagators, but not in the numerators. 
Another natural choice is to consider for ${\mathfrak f}^{\hat{a}\hat{b}\hat{c}}$ a finite dimensional algebra, 
such as $SU(N)$, where $N$ is unrelated to the rank of the gauge group\footnote{Indeed, the color-ordered amplitudes of these theories have recently been discussed in~\cite{Du:2011js}, where the KK and BCJ relations were explicitly shown to hold.}.

In this way one finds a large class of theories whose amplitudes satisfy the KK and BCJ relations. 
While these theories may have little intrinsic interest on their own, we will find a use for them below.

\section{Cubic vertices for Yang-Mills theory}
\label{sec:cubicYM}
\noindent
In the previous section, we encountered a large family of theories which share the algebraic properties of Yang-Mills amplitudes. We will now use these theories to form a basis for the Yang-Mills amplitudes, using only the KK and BCJ relations. The starting point in our analysis is a subset of color-ordered amplitudes, {\it e.g.}
\be
A_{(J)} = A(\sigma_J\{1,\ldots,n-3\},n-2,n-1,n),
\ee
where $\sigma_J$ denotes a permutation of $n-3$ external legs, $J=1,\ldots,(n-3)!$. Recall that all color-ordered amplitudes can be expressed in terms of this subset through the KK and BCJ relations. We now consider the following linear system,
\be
\label{sys}
A_{(J)} = \sum_{I=1}^{(n-3)!} \alpha_I \,\theta_{IJ}.
\ee
Given a non-singular square matrix $\theta_{IJ}$, we can solve for the coefficients $\alpha_I$. In the same manner that we can apply the KK and BCJ relations to the set $A_{(J)}$ to obtain all color-ordered Yang-Mills amplitudes 
(and therefore the color-dressed amplitude $\mathcal{A}$) 
we can now apply those relations to the columns of $\theta_{IJ}$, in order to obtain a color-dressed quantity $\Theta_I$,
\be
\label{standarddecT}
\Theta_I = g^{n-2}  \sum_\sigma  \textrm{Tr}[T^{a_1}T^{a_2}\cdots T^{a_n}] \, \theta_{I} (1,2,\ldots,n),
\ee
such that
\be
\theta_{IJ} =\theta_{I(J)} = \theta_I(\sigma_J\{1,\ldots,n-3\},n-2,n-1,n).
\ee
The full Yang-Mills amplitude is then given by
\be
\label{lincomb}
\mathcal{A} = \sum_{I=1}^{(n-3)!} \alpha_I \,\Theta_I.
\ee
Note that this implies that the $\alpha_I$ are symmetric under permutations of the particles; this important property follows since the amplitudes $\mathcal{A}$ and $\Theta_I$ are permutation symmetric by Bose symmetry. The $\Theta_I$'s form a basis for the Yang-Mills amplitude, and indeed for any amplitude satisfying the KK and BCJ relations. There are only two requirements in this construction: that the elements of the basis $\Theta_I$ satisfy the KK and BCJ relations themselves, and that they are linearly independent. The latter condition corresponds to having a matrix $\theta_{IJ}$ which is non-singular, {\it i.e.} with rank $(n-3)!$.

We now know how to obtain a representation for a Yang-Mills amplitude with manifest color-kinematics duality. Suppose that the basis elements $\Theta_I$ are naturally written in a representation
\be
\label{theta}
\Theta_I = \sum_i \frac{n_{I,i} \, c_i}{D_i},
\ee
where the sum runs over diagrams with cubic vertices, and the numerators $n_{I,i}$ satisfy the same Jacobi identities as the color factors $c_i$. Then it is clear from \eqref{lincomb} that we can write the numerators $n_i$ of a Yang-Mills amplitude in the following way,
\be
\label{BCJnum}
n_i = \sum_{I=1}^{(n-3)!} \alpha_I \,n_{I,i}.
\ee
By linearity, the Yang-Mills numerators $n_i$ also satisfy the Jacobi identities.

We can construct a basis for Yang-Mills amplitudes with manifest color-kinematics duality by using the theories discussed in the last section. Then the $\Theta_I$'s are amplitudes in their own right. We considered theories with a cubic vertex, which contains the structure constants of both the Yang-Mills Lie algebra, and an additional Lie algebra. In particular, we presented two natural examples for the additional Lie algebra: area-preserving diffeomorphisms and $SU(N)$. It is also possible to use combinations of different Lie algebras.

Let us discuss first the application of diffeomorphism algebras. When fictitious momenta are used in the structure constants, $F^{\vp_1 \vp_2 \vp_3}$, as in the example \eqref{FfD}, it is straightforward to construct the basis of auxiliary amplitudes. The matrix $\theta_{IJ}$, whose columns are independent color-ordered amplitudes of $\Theta_I$, has rank $(n-3)!$ when we make a random choice of the fictitious momenta and/or $\Omega^{\mu\nu}$ for each element of the basis. We have checked this explicitly up to 10-points in four dimensions, and up to 9-points in five and six dimensions, and we have checked \eqref{lincomb} using gluon amplitudes obtained from \cite{Bourjaily:2010wh}.

Let us also comment on the use of physical momentum in the auxiliary structure constants, $F^{p_1 p_2 p_3}$. In this case, there can be non-trivial combinations of momentum dependence in the numerators and denominators of the various diagrams. Combinations of this type, which are not obvious when the numerators are written down, lead to the vanishing of amplitudes in the self-dual Yang-Mills theory. In the case of the more general diffeomorphism algebra, such combinations lead to a singular matrix $\theta_{IJ}$ for $n\geq 8$ in four dimensions, if we make a random choice $\Omega^{\mu\nu}$ for each element of the basis $\Theta_I$. However, this choice of auxiliary algebra allows for a special treatment of MHV amplitudes, different from the procedure in \cite{Monteiro:2011pc}. Consider the case where the structure constant is
\be
Y_\lambda(p_1,p_2)= \lambda X(p_1,p_2)  + (1-\lambda)  \bar{X}(p_1,p_2),  \qquad  0<\lambda<1,
\ee
where we use light-cone coordinates $(u=t-z,w=x+iy,\bar{w}=x-iy)$ to define
\be
X(p_1,p_2)= p_{1 w} p_{2 u} - p_{1 u} p_{2 w}, \qquad  \Xb(p_1,p_2)= p_{1 \wb} p_{2 u} - p_{1 u} p_{2 \wb}.
\ee
$X$ corresponds to the self-dual theory, and $\Xb$ to the anti-self-dual theory. We find (up to 10-points) that the rank of $\theta_{IJ}$, using a random value of $\lambda$ for each $I$, is only $n-3$; that is, $\theta_{IJ}$ is singular for $n>5$. However, a solution to the system \eqref{sys} still exists,
\be
\mathcal{A}_{MHV} = \sum_{I=1}^{n-3} \alpha_I \,\Theta_{\lambda_I}\,.
\ee
So the MHV amplitude is part of a certain $(n-3)$-dimensional subspace of the full $(n-3)!$ dimensional 
vector space. 
In~\cite{Monteiro:2011pc}, an explicit color-kinematics dual representation of MHV amplitudes was found in 
light-cone gauge by using one of the negative helicity particles to define the light-cone direction. 
In the expression above, we treat all the external particles democratically. For NMHV amplitudes, we find 
that we can solve the system \eqref{sys} with smaller rank in $n=8$, but we have not explored the general case. 
This should require an extension of the area-preserving diffeomorphism algebra. It would be interesting to 
understand this structure.

The other natural choice for the auxiliary structure constants discussed in the previous section is the Lie algebra of $SU(N)$, $f^{\hat{a}\hat{b}\hat{c}}$, or other simple Lie groups. In order to form a basis, the only restriction is that $N$ must be high enough so that we can find non-vanishing basis amplitudes for $(n-3)!$ different choices of auxiliary indices $\hat{a}$. We have checked that even $SU(2)$ comes short only at 8-points, for which $SU(3)$ is sufficient. Surprisingly, the use of $SU(N)$ Lie algebras transforms the color-kinematics duality into a color-color duality (with unrelated groups).

Notice that the numerators $n_{I,i}$ are constants, independent of the kinematics, for the auxiliary Lie algebras of both $SU(N)$ and the diffeomorphism algebras with fictitious momenta. Then the kinematic dependence of the Yang-Mills amplitude is entirely given in the propagators and in the $(n-3)!$ Lorentz invariant coefficients $\alpha_I$. This may prove useful for applications of our construction (loops, shifts, etc.).

In Appendix~\ref{sec:app}, we illustrate the method described above with an example at five-points.

\section{Amplitude duals}
\label{sec:colordual}
\noindent
The standard color decomposition of an amplitude in gauge theory is
\be
\label{standarddecomp2}
{\mathcal A} = g^{n-2}  \sum_\sigma  \textrm{Tr}[T^{a_1}T^{a_2}\cdots T^{a_n}] \, A(1,2,\ldots,n).
\ee
Given the color-kinematics duality, a natural analogue of this expression was proposed in~\cite{Bern:2011ia},
\be
\label{dualdecomp}
{\mathcal A} = g^{n-2}  \sum_\sigma  \tau_{(1,2,\ldots,n)} \, A^\mathrm{dual}(1,2,\ldots,n).
\ee
The kinematic prefactors $\tau_{(1,2,\ldots,n)}$ are the analogues of color traces, and $A^\mathrm{dual}$ is a dual amplitude constructed from the color-ordered gauge theory amplitude $A$ by replacing all kinematic numerators with color factors, {\it i.e.} $n_i \to c_i$. It was shown in~\cite{Bern:2011ia} how to write 
prefactors $\tau_{(1,2,\ldots,n)}$ in terms of the numerators $n_i$. However, the resulting expressions are 
a bit complicated and perhaps not very illuminating.

We can now use our cubic vertex formalism to write kinematic prefactors $\tau_{(1,2,\ldots,n)}$. Consider the basis elements $\Theta_I$,
\be
\Theta_I = \sum_i \frac{n_{I,i} \, c_i}{D_i},
\ee
with the numerators $n_{I,i}$ satisfying the same Jacobi identities as the color factors $c_i$. Suppose that the auxiliary Lie algebra used to construct the numerators is the Lie algebra of $SU(N)$, as discussed in previous sections. In this case, the numerators $n_{I,i}$ are also color factors, with particle labels $\hat{a}$, while the Yang-Mills color factors $c_i$ have particle labels $a$. Therefore, we can take the standard decomposition,
\be
\Theta_I = \sum_\sigma  \textrm{Tr}[T^{a_1}T^{a_2}\cdots T^{a_n}] \, \theta_I(1,2,\ldots,n),
\ee
and invert the roles of the two Lie algebras,
\be
\Theta_I = \sum_\sigma  \textrm{Tr}[T^{\hat{a}_1}T^{\hat{a}_2}\cdots T^{\hat{a}_n}] \, A^\mathrm{dual}(1,2,\ldots,n).
\ee
The indices in the color matrices are now the auxiliary color indices, which differ for each value of $I$. Our result,
\be
\label{lincomb2}
\mathcal{A} = \sum_{I=1}^{(n-3)!} \alpha_I \,\Theta_I,
\ee
and the expression~\eqref{standarddecomp2} lead to
\be
\label{tautraces}
\tau_{(1,2,\ldots,n)} = \sum_{I=1}^{(n-3)!} \alpha_I \, \textrm{Tr}[T^{\hat{a}_1}T^{\hat{a}_2}\cdots T^{\hat{a}_n}].
\ee
The cyclic properties of $\tau_{(1,2,\ldots,n)}$ are inherited from the color traces, and from the fact that the coefficients $\alpha_I$ are invariant under permutations of the external particles.

\section{Gravity amplitudes}
\label{sec:gravity}
\noindent

Gravity amplitudes can be obtained as the ``square" of gauge theory amplitudes at the level of numerators~\cite{Bern:2008qj,Bern:2010yg}, as detailed in eq. (\ref{square}). The fact that we can write a gauge theory amplitude in the dual form \eqref{dualdecomp} 
then implies that~\cite{Bern:2011ia}
\be
\label{gravtau}
{\mathcal M} = i \left( \frac{\kappa}{2} \right)^{n-2}  \sum_\sigma  \tau_{(1,2,\ldots,n)} \, A(1,2,\ldots,n),
\ee
where $A$ denotes a color-ordered gauge theory amplitude.\footnote{We could have considered $\tilde{A}$, to denote that there were two distinct gauge theories, as in (\ref{square}).} Comparing with the expression \eqref{standarddecomp2}, we can see that the kinematic prefactors $\tau_{(1,2,\ldots,n)}$ play the same role for gravity amplitudes as the color traces play for gauge theory amplitudes. It is remarkable that $SU(N)$ color traces can play a role in gravity amplitudes, through our expression \eqref{tautraces}. We can also rearrange the sums and write a basis for gravity amplitudes analogous to our result \eqref{lincomb2},
\be
{\mathcal M} = \sum_{I=1}^{(n-3)!} \alpha_I' \,\mathcal{A}_I.
\ee
The basis elements are now color-dressed gauge theory amplitudes with specific choices of labels $\hat{a}$,
\be
\mathcal{A}_I  = \sum_\sigma  \textrm{Tr}[T^{\hat{a}_1}T^{\hat{a}_2}\cdots T^{\hat{a}_n}] \, A(1,2,\ldots,n),
\ee
and we have defined $\alpha_I' = i \left( \kappa/2 \right)^{n-2} \alpha_I$.

This suggests the following physical interpretation for the color-kinematics duality. It is an old idea to try to view gravity as a gauge theory of an infinite-dimensional group, the group of diffeomorphisms. 
However, in a given scattering process there is only a finite number of particles with definite momenta. 
Let us recall the use of area-preserving diffeomorphisms, reviewed in Section~\ref{sec:cubicScalar}. These were seen in~\cite{Monteiro:2011pc} to directly underlie the color-kinematics duality for some four-dimensional amplitudes. A set of generators $L_{p_a}$ of the Lie algebra, each associated with the momentum of a particle, leads to a finite-dimensional sub-algebra when the momentum is conserved, $\sum_a p_a=0$. The dimension of the sub-algebra, which corresponds to all distinct momenta that can run through propagators and external legs in a Feynman diagram expansion, grows with the number of particles $n$. Considering the Dynkin classification of infinite families of finite-dimensional simple Lie algebras, it is perhaps not surprising that the role of the general kinematic algebra can be played by the Lie algebra of $SU(N)$, where $N$ should just be sufficiently large (the minimum required to construct a basis of auxiliary amplitudes $\Theta_I$). For the set of all amplitudes, with arbitrary number of legs, an infinite-dimensional group, perhaps along the lines of the Bondi-van der Burg-Metzner-Sachs group~\cite{group}, is required.

\section{Conclusions}
\label{sec:concl}

We have presented a systematic way to express tree-level gauge theory amplitudes in a manifest 
color-kinematics dual representation. The KK and BCJ relations play crucial roles in this construction, 
allowing us to write the gauge theory amplitude using a basis of auxiliary amplitudes with 
convenient properties. The freedom in choosing the kinematic numerators is translated into the freedom 
in choosing this basis. The duality between color and kinematics is made manifest by using auxiliary 
amplitudes whose cubic vertices are dressed with the structure constants of both the gauge theory 
Lie algebra, $f^{abc}$, and an additional ``kinematic" Lie algebra, ${\mathfrak f}^{\hat{a}\hat{b}\hat{c}}$, 
a procedure inspired by~\cite{Monteiro:2011pc}. In all of this lies an inspiration coming from the
MHV-sector of the theory. The intriguing connection between MHV amplitudes and matrix theory 
\cite{Heckman:2011qt} may also lead to insight here.  

For gravity amplitudes, we have determined objects which play roles analogous to color traces in gauge 
theory amplitudes, expressing them in terms of a basis of actual color traces. We believe this surprising result 
has the potential to further our understanding of gravity amplitudes, and to simplify their calculations.

\acknowledgments

We thank Simon Badger, Zvi Bern, Henrik Johannson and Thomas S\o ndergaard for useful discussions. NEJBB is grateful for hospitality at the IHES while this work was being finished.

\appendix

\section{Example at five-points}
\label{sec:app}

In this appendix, we illustrate our method to construct kinematic numerators satisfying Jacobi-like identities with an explicit example. For simplicity, we consider 5-point amplitudes and use auxiliary structure constants from the Lie algebra of $SU(2)$. Recalling \eqref{eq:3pt}, we have now ${\mathfrak f}^{\hat{a}\hat{b}\hat{c}} \to \epsilon^{\hat{a}\hat{b}\hat{c}}$, where $\epsilon^{\hat{a}\hat{b}\hat{c}}$ is the Levi-Civita symbol. The linear system \eqref{sys} is two-dimensional, and we can choose the following color-ordered amplitudes as the independent set:
\bea
A(1,2,3,4,5)  = \alpha_1 \,\theta_1 (1,2,3,4,5) + \alpha_2 \,\theta_2 (1,2,3,4,5), \nonumber \\
A(2,1,3,4,5)  = \alpha_1 \,\theta_1 (2,1,3,4,5) + \alpha_2 \,\theta_2 (2,1,3,4,5).
\eea
For concreteness, we choose the auxiliary color indices to be
\bea
\{\hat{a}_1,\hat{a}_2,\hat{a}_3,\hat{a}_4,\hat{a}_5\}=\{1,2,3,3,3\} \quad \to \quad \theta_1, \nonumber \\
\{\hat{a}_1,\hat{a}_2,\hat{a}_3,\hat{a}_4,\hat{a}_5\}=\{1,1,1,2,3\} \quad \to \quad \theta_2.
\eea
The system is then quite simple:
\bea
A(1,2,3,4,5)  = -\frac{\alpha_1}{s_{15}\,s_{23}} -\frac{\alpha_2}{s_{15}\,s_{34}} , \nonumber \\
A(2,1,3,4,5)  = -\frac{\alpha_1}{s_{13}\,s_{25}} -\frac{\alpha_2}{s_{25}\,s_{34}} .
\eea
This can be inverted as
\begin{align}
\alpha_1  &= \frac{s_{13}\,s_{23}}{s_{13}+s_{23}} ( -s_{15}A(1,2,3,4,5) +s_{25}A(2,1,3,4,5)) , \nonumber \\
\alpha_2  &= \frac{s_{34}}{s_{13}+s_{23}} ( s_{15}\,s_{23}A(1,2,3,4,5) +s_{13}\,s_{25}A(2,1,3,4,5)) .
\label{app:inv}
\end{align}
Using the expressions \eqref{lincomb} and \eqref{BCJnum}, we can write the amplitude in the BCJ representation \eqref{bcjrep}:
\begin{align}
{\mathcal A}  &= g^3\Bigg( - \frac{\alpha_2}{s_{24}\,s_{35}}\, c(24,1,35) +\frac{\alpha_2}{s_{25}\,s_{34}}\, c(25,1,34)
+ \frac{\alpha_2}{s_{15}\,s_{34}}\, c(15,2,34) - \frac{\alpha_2}{s_{14}\,s_{35}}\, c(14,2,35)  \nonumber \\
&\qquad -\frac{\alpha_1+\alpha_2}{s_{14}\,s_{25}}\, c(14,3,25) - \frac{\alpha_1-\alpha_2}{s_{15}\,s_{24}}\, c(15,3,24)  
- \frac{\alpha_1}{s_{15}\,s_{23}}\, c(15,4,23) - \frac{\alpha_1}{s_{13}\,s_{25}}\, c(13,4,25) \nonumber \\
&\qquad - \frac{\alpha_1}{s_{13}\,s_{24}}\, c(13,5,24) - \frac{\alpha_1}{s_{14}\,s_{23}}\, c(14,5,23) \Bigg),
\end{align}
where we denote the color factors as, for instance, $c(24,1,35) = f^{a_2 a_4 b_1} \, f^{b_1 a_1 b_2} \, f^{b_2 a_3 a_5}$.

By construction, the kinematic numerators satisfy the same Jacobi identities as the color factors, {\it e.g.}
\be
c(14,2,35) - c(14,3,25) + c(14,5,23) = 0    \quad \longrightarrow \quad -\alpha_2+(\alpha_1+\alpha_2) - \alpha_1=0.
\ee

The gravity amplitude obtained through the squaring relation \eqref{square} is
\begin{align}
M  &= i\, \frac{\kappa^3}{8} \Bigg( \frac{\alpha_2^2}{s_{24}\,s_{35}} +\frac{\alpha_2^2}{s_{25}\,s_{34}}
+ \frac{\alpha_2^2}{s_{15}\,s_{34}} + \frac{\alpha_2^2}{s_{14}\,s_{35}} 
+\frac{(\alpha_1+\alpha_2)^2}{s_{14}\,s_{25}}+ \frac{(\alpha_1-\alpha_2)^2}{s_{15}\,s_{24}} \nonumber \\
&\qquad \qquad  + \frac{\alpha_1^2}{s_{15}\,s_{23}} + \frac{\alpha_1^2}{s_{13}\,s_{25}} 
+ \frac{\alpha_1^2}{s_{13}\,s_{24}} + \frac{\alpha_1^2}{s_{14}\,s_{23}}  \Bigg).
\end{align}
%


\end{document}